\documentclass[12pt]{article}

\input epsf
\usepackage{amssymb}
\usepackage[dvips]{graphicx}
\usepackage{amscd}
\usepackage{amsmath}

\def\bb#1{{\tt hep-th/#1}}

\def\stackrel#1#2{\mathrel{\mathop{#2}\limits^{#1}}}

\long\def\symbolfootnote[#1]#2{\begingroup%
   \def\thefootnote{\fnsymbol{footnote}}\footnote[#1]{#2}\endgroup} 

\def\dj{\hbox{d\kern-0.347em \vrule width 0.3em height 1.252ex depth
-1.21ex \kern 0.051em}}

\numberwithin{equation}{section}

\renewcommand{\appendix}{
	\renewcommand{\section}{
		\newpage\thispagestyle{plain}
		\secdef\Appendix\sAppendix}
	\setcounter{section}{0}
	\renewcommand{\thesection}{\Alph{section}}
}

\begin{document}

\setlength{\oddsidemargin}{0cm}
\setlength{\baselineskip}{7mm}


\thispagestyle{empty}
\setcounter{page}{0}

\begin{flushright}
CERN-PH-TH/033-2005   \\
{\tt hep-th/0503016}
\end{flushright}

\vspace*{1cm}

\begin{center}
{\bf \LARGE On nonanticommutative  $\boldsymbol{\mathcal{N}=2}$ sigma-models 
}

\vspace*{0.5cm}

{\bf \LARGE in two dimensions}

\vspace*{1cm}

Luis \'Alvarez-Gaum\'e$\,^{\rm a,\,\,}$\symbolfootnote[1]{\tt 
Luis.Alvarez-Gaume@cern.ch} 
and Miguel A. V\'azquez-Mozo$\,^{\rm b,\,\, c,\,\, }$\symbolfootnote[3]{\tt 
Miguel.Vazquez-Mozo@cern.ch}

\end{center}

\vspace*{0.25cm}

\begin{quote}
$^{\rm a}$ {\sl Physics Department, Theory Division, CERN, CH-1211
Geneva 23, Switzerland}


$^{\rm b}$ {\sl F\'{\i}sica Te\'orica, Universidad de Salamanca, 
Plaza de la Merced s/n, \\ E-37008 Salamanca, Spain}


$^{\rm c}$ {\sl Instituto Universitario de F\'{\i}sica Fundamental y
Matem\'aticas (IUFFyM), \\ Universidad de Salamanca, Salamanca, Spain}

\end{quote}

\vspace*{1cm}

\centerline{\bf \large Abstract}

\vspace*{0.5cm}

\noindent
We study nonanticommutative deformations of $\mathcal{N}=2$
two-dimensional Euclidean sigma models. We find that these theories
are described by simple deformations of Zumino's Lagrangian and the
holomorphic superpotential. Geometrically, this deformation can be
interpreted as a fuzziness in target space controlled by the
vacuum expectation value of the auxiliary field. In the case of
nonanticommutative deformations preserving Euclidean invariance, we
find that a continuation of the deformed supersymmetry algebra to
Lorentzian signature leads to a rather intriguing central extension of
the ordinary (2,2) superalgebra.

\newpage

\section{Introduction}

The study of deformations of quantum field theories has been the
subject of renewed interest since the realization of its relevance in
string theory. A large part of these work has been devoted to
noncommutative theories where the commutator of the coordinates gets
deformed by a central term. Quantum fields defined on noncommutative
spaces present a plethora of unexpected properties \cite{noncommutative}.

More recently it has been realized that a new class of theories with
reduced supersymmetry can be constructed by deforming the 
anticommutator of the fermionic coordinates in superspace. In Ref. 
\cite{seiberg} $\mathcal{N}=1$ supersymmetric theories in four dimensions were 
deformed by
\begin{eqnarray}
\{\theta^{a},\theta^{b}\}=C^{ab}, \hspace*{1cm} C^{ab}\in\mathbb{C},
\label{nonanti}
\end{eqnarray}
while all other anticommutators are equal to zero.  On general
grounds, a Clifford deformation \cite{H} of the Grassmann algebra of
fermionic coordinates can only be consistently carried out in
Euclidean signature where $\theta^{\pm}$ and $\overline{\theta}{}^{\pm}$
are independent \cite{Klemm}. Eq. (\ref{nonanti}) induces a
deformation in the $\mathcal{N}=1$ supersymmetry algebra such that
only one half of the original supercharges remain a symmetry of the
deformed theory \cite{seiberg,Ferrara,terashima,araki}. This kind of
theories emerge also as an effective low energy description of certain
superstring theories in constant graviphoton backgrounds \cite{low}. In
this set-up the Euclidean signature is forced by the requirement that
the background field has vanishing back reaction on the metric.

In the case of the nonanticommutative versions of $\mathcal{N}=1$
super-Yang-Mills theories and the Wess-Zumino model the deformation
results in the addition of a finite number of higher dimensional
operators to the undeformed Lagrangian \cite{seiberg}. This implies in
particular that the deformed theory preserves locality, unlike the
case of noncommutative field theories where the Lagrangian contains
and infinite number of terms with an arbitrary high number of
derivatives. This type of theories are surprisingly renormalizable
in spite of the presence of nominally higher dimensional terms in the
Lagrangian \cite{renorm,BFR}. In the case of theories with more
supersymmetries the situation is richer. For example,
nonanticommutative singlet deformations of four-dimensional
$\mathcal{N}=2$ theories \cite{ivanov} give rise to a non-polynomial
deformed superpotential \cite{ferrara2}. Instantons in nonanticommutative
theories have been also a subject of interest \cite{instanton}.

Two-dimensional supersymmetric sigma models with extended
supersymmetry are a particularly interesting class of theories to
study nonanticommutative deformations.  In \cite{ck,c} the deformation
of (2,2) two-dimensional sigma-models was considered and it was found
that the deformed Lagrangian can be written as an infinite series
in $\det{C}$ (see \cite{azorkina} for its four-dimensional
analog). Each term, however, contains at most two
derivatives, so one may ask whether there is a possibility of
rewriting a Lagrangian containing a finite number of terms with
deformed (field-dependent) couplings.

In this paper we have a closer look at the construction of such
nonanticommutative $\mathcal{N}=2$ sigma-models in two dimensions.  In
particular we consider a type of nonanticommutative deformation that
preserve two-dimensional Euclidean invariance. We find that, in this
particular case, the supersymmetry algebra is
deformed by a central extension depending on the Casimir operator 
$P_{+}P_{-}$.

Despite the low number of dimensions we work in, and the fact that
chirality and charge conjugation are compatible in two-dimensions,
it seems not possible to find a ``two-dimensional loophole'' of 
\cite{low} (specially the last reference) which would have allowed
us to work with Lorentzian world-sheets. We can, nevertheless, press
on and consider the deformed, centrally extended supersymmetry algebra
in Minkowski space-time. In this case we have a central extension of
the two-dimensional $\mathcal{N}=2$ supersymmetry algebra with a
central extension compatible with Haag-\L opusza\'nski-Sohnius theorem
\cite{hls}, where the central charges depend on kinematical 
invariants like the mass of the state. The interesting part is that
the superalgebra implies a bound for the masses of the states in the
theory.  We are not aware of any supersymmetric field theory realizing
such an algebra, but its existence would be quite interesting since
it automatically includes an ultraviolet cutoff.

We also show that the infinite series found in \cite{ck,c} can be
resummed in a compact expression which can be written as a the
standard Zumino's Lagrangian \cite{zumino} with deformed
K\"ahler potential and superpotential, plus a finite number of higher
dimensional terms with deformed (field-dependent) couplings.
Interestingly, the deformation of the K\"ahler potential
$K(z,\overline{z})$ and the holomorphic superpotential
$\mathcal{W}(z)$ has the physical interpretation of a smearing in the
holomorphic coordinates of the target manifold $\varphi^{i}$
controlled by the auxiliary field $F^{i}$. Although we study the case
of a particular Euclidean-invariant deformation $C^{\pm\pm}=0$,
$C^{\pm\mp}=1/M$, the fact that the deformed Lagrangian depends on
$C^{ab}$ only through $\det{C}$ \cite{ck,c} implies that our results
remain valid for other deformation matrices.

The plan of the paper is as follows: in Section 2 we briefly review 
the Weyl map formalism for nonanticommutative theories. Section 3
is devoted to summarizing some general aspects of the deformation 
of two-dimensional (2,2) supersymmetric theories. Using the formalism
of Section 2 we calculate in Section 3 the deformed superpotential and
K\"ahler potential, and discuss the physical interpretation of the 
result. Section 4 deals with the discussion of the classical structure
of vacua for the deformed theory. Finally in Section 5 we summarize our
conclusions. 

To make the paper more readable some technical details have been
postponed to the Appendices.  In Appendix A aspects of Fourier
transforms of functions of anticommuting variables are
reviewed. Appendix B details some of the calculation of Section 3. 
In Appendix C we include a discussion on the representation of
the deformed supersymmetry algebra.

\section{The Weyl map}
\label{weyl}

Noncommutative field theories can be conveniently formulated using the
Weyl map which defines a homomorphism between the noncommutative
algebra of functions and an algebra of operators. To begin with let us
consider a superspace $\mathbb{R}^{m|2n}$ with anticommuting
coordinates $\theta^{1},\ldots,\theta^{2n}$. The deformation
(\ref{def}) amounts then to deforming the $\mathbb{Z}_{2}$-graded
algebra of functions defined on it. Following the bosonic case the
idea is to find a map between this deformed graded algebra of
functions and a graded algebra of operators. Consequently, a set of
degree one operators $\widehat{Q}^{a}$ ($a,b=1,\ldots,2n$) is then
introduced satisfying the deformed anticommutation relations
\begin{eqnarray}
\{\widehat{Q}^{a},\widehat{Q}^{b}\}=C^{ab}.
\end{eqnarray}
In terms of them any superspace function $f(\theta)$ has associated its
Weyl transform (or Weyl symbol) through the definition 
\begin{eqnarray}
\widehat{f}\equiv (-1)^{n}\int d^{2n}\eta\,e^{-\eta_{a}
\widehat{Q}^{a}}\widetilde{f}(\eta),
\label{Q}
\end{eqnarray}
where $\widetilde{f}(\eta)$ is the Fourier transform of $f(\theta)$
(see Appendix A for notation and definitions). This definition of the
symbol associated to a superspace function $f(\theta)$ will be very
important in the calculation of the action of the deformed Lagrangian
in the next section.

The set of Weyl symbols form another $\mathbb{Z}_{2}$-graded algebra with
${\rm deg}(\widehat{f})={\rm deg}(f)$.  Actually, the map (\ref{Q})
defines a homomorphism between the two algebras in which the product
of two symbols $\widehat{f}$, $\widehat{g}$ is associated with the
noncommutative product of the corresponding functions
\begin{eqnarray}
\widehat{f}\,\,\widehat{g}=\int d^{2n}\eta 
\,e^{-\eta_{a}\widehat{Q}^{a}} \widetilde{
(f\star g)}(\eta),
\label{productWeyl}
\end{eqnarray}
where the star-product is defined by\footnote{We follow the conventions
of Ref. \cite{seiberg} and define right derivations as 
\begin{eqnarray}
f(\theta)
\stackrel{\longleftarrow}{{\partial\over \partial\theta^{a}}}
=(-1)^{{\rm deg}(f)}
\stackrel{\longrightarrow}{{\partial\over \partial\theta^{a}}}
f(\theta)
\end{eqnarray}
}
\begin{eqnarray}
f(\theta)\star g(\theta) = f(\theta) \exp\left({-{1\over 2}C^{ab}
\stackrel{\longleftarrow}{{\partial\over \partial\theta^{a}}}
\stackrel{\longrightarrow}{{\partial\over \partial\theta^{b}}}}\right)
g(\theta).
\label{star1}
\end{eqnarray}

The function-symbol map (\ref{Q}) can be also conveniently expressed
using an operator $\widehat{\Xi} (\theta)$ defined by
\begin{eqnarray}
\widehat{\Xi}(\theta)=(-1)^{n}\int d^{2n}\eta\,e^{-\eta_{a}(\widehat{Q}^{a}-
\theta^{a})}.
\label{xi}
\end{eqnarray}
A trivial manipulation of (\ref{Q}) shows that $\widehat{f}$ can be written
as the integral
\begin{eqnarray}
\widehat{f}=\int d^{2n}\theta\,\,\widehat{\Xi}(\theta)\,f(\theta).
\label{inverse}
\end{eqnarray}
Using Eq. (\ref{xi}) it is easy to construct the inverse of the Weyl map
by defining a supertrace operation on the algebra of symbols with the
following properties
\begin{eqnarray}
{\rm str}\,{\bf 1} &=& 0, \nonumber \\ 
{\rm str}\, 
\left[\widehat{Q}^{a_{1}}\ldots\widehat{Q}^{a_{k}}\right]&=&0, 
\hspace*{1cm} k=1,\ldots, 2n-1 \nonumber \\
{\rm str}\left[\widehat{Q}^{1}\ldots\widehat{Q}^{2n}\right]&=&1.
\end{eqnarray}
Then, given a symbol $\widehat{f}$ the function $f(\theta)$ associated
with it by the map (\ref{Q}) is given by
\begin{eqnarray}
f(\theta)={\rm str\,}\left[\widehat{f}\,\,\widehat{\Xi}(\theta)\right].
\end{eqnarray}

For later use it is interesting to particularize some of the formulae
to the case $n=1$ with fermionic coordinates $\theta^{\pm}$. In this
case the symbol $\widehat{f}$ associated with the function $f(\theta^{\pm})$
is defined in terms of its Fourier transform $\widetilde{f}(\eta_{\pm})$ as
\begin{eqnarray}
\widehat{f}=-\int d^{2}\eta\,e^{-(\eta_{+}\widehat{Q}^{+}+
\eta_{-}\widehat{Q}^{-})}\widetilde{f}(\eta_{\pm}),
\label{symbolpm}
\end{eqnarray}
whereas the inverse map is constructed as in Eq. (\ref{inverse}) with 
$\widehat{\Xi}(\theta^{\pm})$ given by
\begin{eqnarray}
\widehat{\Xi}(\theta^{\pm})=-\int d^{2}\eta\,\exp{\left[\eta_{+}
\left(\widehat{Q}^{+}-\theta^{+}\right)
+\eta_{-}\left(\widehat{Q}^{-}-\theta^{-}\right)\right]}.
\end{eqnarray}

The Weyl formalism gives automatically a prescription for the
definition of functions of the superfields defined on the deformed
superspace. Given a function $F(\Phi)$ of a superfield $\Phi(\theta)$ 
the corresponding operator $F(\widehat{\Phi})$ 
is defined by
\begin{eqnarray}
F(\widehat{\Phi})\equiv \sum_{n=0}^{\infty}{1\over n!}
F^{(n)}(0)\widehat{\Phi}^{n}.
\end{eqnarray}
The deformed function $F(\Phi)_{\star}$ is then obtained by applying the 
inverse Weyl map
\begin{eqnarray}
F(\Phi)_{\star}&=&{\rm str\,}
\left[\widehat{\Xi}(\theta)\,F(\widehat{\Phi})\right]
=\sum_{n=0}^{\infty}{1\over n!}F^{(n)}(0)\,\Phi(\theta)\star
\stackrel{(n)}{\ldots}\star\Phi(\theta).
\end{eqnarray}
This recovers the usual prescription for the definition of functions
of superfields in nonanticommutative superspaces (see, for example,
\cite{ck,c}).

\section{Nonanticommutative theories in two dimensions}

In the following we focus our analysis to the case of two-dimensional
(2,2) superspace $\mathbb{R}^{2|4}$ with coordinates $x^{\pm}$, 
$\theta^{\pm}$ and $\overline{\theta}^{\,\pm}$.  Although
the theory is defined in Euclidean space we use Minkowskian notation
throughout and consider the Euclidean invariant deformation
\begin{eqnarray}
\{\theta^{+},\theta^{-}\}={1\over M}, \hspace*{1cm} (\theta^{\pm})^{2}=0,
\label{def}
\end{eqnarray}
with $M\in\mathbb{R}$ the characteristic energy scale of the
deformation and all the anticommutators of antiholomorphic coordinates
unchanged. Since we are working in Euclidean space, it is very
important to keep in mind that bars do not denote complex conjugation,
$\overline{\theta} {}^{\pm}\neq (\theta^{\pm})^{*}$. This theory can
be obtained by dimensional reduction from the corresponding
$\mathcal{N}=1/2$ theory in four dimensions \cite{c} by taking
$C^{11}=C^{22}=0$, $C^{12}=C^{21}=1/M$.

\subsection{The deformed superalgebra}

In order to make the chiral structure of the theory explicit it is
convenient to introduce chiral coordinates $y^{\pm}$ given by
\begin{eqnarray}
y^{\pm}=x^{\pm}-i\theta^{\pm}\overline{\theta}^{\,\pm},
\end{eqnarray}
which satisfy $[y^{+},y^{-}]=0$ provided that
\begin{eqnarray}
[x^{\pm},\theta^{\mp}]=-{i\over M}\overline{\theta}^{\,\pm}, 
\hspace*{1cm} [x^{+},x^{-}]={1\over M}\overline{\theta}^{\,+}
\overline{\theta}^{-},
\label{commutation}
\end{eqnarray}
with all other commutators involving $x^{\pm}$ equal to
zero. Conditions (\ref{commutation}) also imply that $y^{\pm}$ commute
with the holomorphic fermionic coordinates $\theta^{\pm}$. This fact
is actually crucial to apply the formalism of the previous section,
since it allows to consider $y^{\pm}$ as spectator coordinates in the
Weyl map.

In terms of $y^{\pm}$ the supercovariant derivatives are defined in
the usual way \cite{seiberg,c}
\begin{eqnarray}
D_{\pm}= {\partial\over \partial\theta^{\pm}}-2i\overline{\theta}^{\,\pm}
{\partial\over \partial y^{\pm}}, \hspace{2cm}
\overline{D}_{\pm}= 
-{\partial\over \partial\overline{\theta}^{\,\pm}},
\end{eqnarray}
and similarly for the supercharges $Q_{\pm}$, $\overline{Q}_{\pm}$ 
\begin{eqnarray}
Q_{\pm}= {\partial\over \partial\theta^{\pm}}, \hspace{2cm}
\overline{Q}_{\pm}= 
-{\partial\over \partial\overline{\theta}^{\,\pm}}-2i{\theta}^{\pm}
{\partial\over \partial y^{\pm}}.
\end{eqnarray}
The deformation (\ref{def}) implies that the algebra of supercharges 
and supercovariant derivatives remain unchanged except for the only 
anticommutator
\begin{eqnarray}
\{\overline{Q}_{+},\overline{Q}_{-}\}=-{4\over M}{\partial^2\over \partial y^+
\partial y^{-}}.
\label{algebra}
\end{eqnarray}
As a result the original (2,2) supersymmetry is broken down to the
supersymmetries generated by $Q_{\pm}$ \cite{seiberg}.

It is rather remarkable that, when expressed in momentum space, the
particular deformation of the algebra (\ref{algebra}) can be written
as
\begin{eqnarray}
\{\overline{Q}_{+},\overline{Q}_{-}\}={4\over M}P_{+}P_{-}.
\label{p2}
\end{eqnarray}
Actually, $P_{+}P_{-}$ is a Casimir operator of the two-dimensional
Euclidean group and therefore the deformed algebra is a central
extension of the (2,2) superalgebra.  It is important to stress that
this central extension (\ref{p2}) only arises for the deformation
(\ref{def}) that preserves Euclidean invariance. Similarly, in
dimensions higher than two any non-vanishing $C^{ab}$ breaks the full
Euclidean invariance of the theory. As a consequence the deformation
of the supersymmetry algebra is not a central extension, unlike the
case studied here.

Although we are forced to work in Euclidean space, it is interesting
to look at the central extension (\ref{p2}) from a Minkowskian point
of view. In principle, such a central extension of the two-dimensional
(2,2) supersymmetry algebra is allowed by the Haag-\L
opusza\'nski-Sohnius theorem \cite{hls}.  Interestingly, following the
arguments of Ref. \cite{WittenOlive} shows that such a central
extension implies an upper bound in the spectrum of eigenvalues of the
mass of the states, given by the Casimir operator $P_{+}P_{-}$.  This
bound is determined by the energy scale of the deformation $M$,
\begin{eqnarray}
P_{+}P_{-}\leq {M^2\over 4}.
\end{eqnarray}
In Appendix C we have given a detailed calculation leading to this
result. It is still to be seen, however, whether this deformed algebra
can be somehow realized in a Quantum Field Theory in Minkowski space-time.

\subsection{Chiral and antichiral superfields}

To fix notation, in the following we will outline the construction of 
chiral and antichiral superfields done in Refs. \cite{seiberg,ck,c}.
Because of the nonanticommutativity of the fermionic coordinates
$\theta^{\pm}$, an ordering prescription is required in the definition
and multiplication of superfields. Weyl (symmetric) ordering is
implemented by introducing the nonanticommutative star-product of
Eq. (\ref{star1}) \cite{seiberg}, which for the particular 
deformation (\ref{def}) reads
\begin{eqnarray}
f(\theta^{\pm})\star g(\theta^{\pm})&=& f(\theta^{\pm})\,{\rm exp}
\left[-{1\over 2M}\left(
\stackrel{\longleftarrow}{{\partial\over \partial\theta^{+}}}
\stackrel{\longrightarrow}{{\partial\over \partial\theta^{-}}}+
\stackrel{\longleftarrow}{{\partial\over \partial\theta^{-}}}
\stackrel{\longrightarrow}{{\partial\over \partial\theta^{+}}}
\right)\right]\,g(\theta^{\pm}) \\
&=& fg-{1\over 2M}\left({\partial f\over \partial\theta^{+}}
{\partial g\over \partial\theta^{-}}+
{\partial f\over \partial\theta^{-}}{\partial g\over \partial\theta^{+}}
\right)
+{1\over 4M^2}\left({\partial^2 f\over \partial\theta^{+}\partial\theta^{-}}
\right)\left({\partial^2 g\over \partial\theta^{+}\partial\theta^{-}}\right),
\nonumber 
\label{star}
\end{eqnarray}
for any two functions $f(\theta^{\pm})$, $g(\theta^{\pm})$ of degree zero. 
The difference between the star-product and the ordinary product is 
a total derivative in $\theta^{\pm}$.

Given that $D_{\pm}$, $\overline{D}_{\pm}$ anticommute with one
another, chiral and antichiral superfields can be defined in the usual
way by\footnote{Twisted chiral or antichiral superfields can also be
defined \cite{c}. Here, however, we only deal with chiral and
antichiral fields}
\begin{eqnarray}
\overline{D}_{\pm}\Phi=0 \hspace*{0.5cm} \mbox{(chiral)}, 
\hspace*{2cm} D_{\pm}\overline{\Phi}=0 \hspace*{0.5cm} \mbox{(antichiral).}
\end{eqnarray}
The constraint for chiral superfields can be easily solved in terms
of $y^{\pm}$ as
\begin{eqnarray}
\Phi(y^{\pm},\theta^{\pm})=\varphi(y^{\pm})+\theta^{+}\psi_{+}(y^{\pm})
+\theta^{-}\psi_{-}(y^{\pm})+\theta^{+}\theta^{-}F(y^{\pm}).
\label{chiral}
\end{eqnarray}
This expression is automatically Weyl ordered, since
$\theta^{+}\theta^{-}={1\over 2}(\theta^{+}\theta^{-}-\theta^{-}\theta^{+})$.

Similarly, antichiral superfields $\overline{\Phi}$ are functions only of 
$\overline{\theta}^{\pm}$ and the antichiral coordinates 
\begin{eqnarray}
\overline{y}^{\pm}=y^{\pm}+2i\theta^{\pm}\overline{\theta}{}^{\,\pm}.
\end{eqnarray}
Using Eq. (\ref{commutation}) one sees that the $\overline{y}{}^{\,\pm}$ do
not commute among themselves or with $\theta^{\pm}$. However the
component fields can be expanded around the commuting coordinates
$y^{\pm}$.  As we will see below, it is convenient to write the
antichiral superfield $\overline{\Phi}$ as a (0,2) superfield whose
component fields are themselves (2,0) superfields
\cite{seiberg,ck,c}:
\begin{eqnarray}
\overline{\Phi}(\overline{y}^{\pm},\overline{\theta}^{\,\pm})
&=& \overline{\varphi}(\overline{y}^{\,\pm})+\overline{\theta}^{\,+}
\overline{\psi}_{+}(\overline{y}^{\,\pm})
+\overline{\theta}^{\,-}\overline{\psi}_{-}(\overline{y}^{\,\pm})+
\overline{\theta}^{+}\overline{\theta}^{\,-}F(\overline{y}^{\,\pm}) 
\label{antichiral} \\
&=& \overline{\varphi}(y^{\pm})
+\overline{\theta}^{\,+}\Big[\overline{\psi}_{+}(y^{\pm})
-2i\theta^{+}\partial_{+}\overline{\varphi}(y^{\pm})\Big]
+\overline{\theta}^{\,-}\left[\overline{\psi}_{-}(y^{\pm})
-2i\theta^{-}\partial_{-}\overline{\varphi}(y^{\pm})\right] \nonumber \\
&+&\overline{\theta}^{\,+}\overline{\theta}^{\,-}\Big[\overline{F}(y^{\pm}) 
+ 2i\theta^{+}\partial_{+}
\overline{\psi}_{-}(y^{\pm})-2i\theta^{-}\partial_{-}
\overline{\psi}_{+}(y^{\pm})
+4\theta^{+}\theta^{-}\partial_{+}\partial_{-}\overline{\varphi}(y^{\pm})\Big].
\nonumber
\end{eqnarray}
Here and in the following $\partial_{\pm}$ indicates derivatives with
respect to $y^{\pm}$. To simplify the notation, from now on we
will not indicate explicitly the dependence of the superfields on the
chiral coordinates $y^{\pm}$ whenever there is no risk of ambiguity.

The unbroken supersymmetries generated by $Q_{\pm}$ act by shifting
the holomorphic fermionic coordinates, $\theta^{\pm}\rightarrow
\theta^{\pm}+\varepsilon^{\pm}$ at fixed $y^{\pm}$
and $\overline{\theta}^{\,\pm}$. The
transformation of the component of a chiral superfield under
$\varepsilon^{+}Q_{+}+\varepsilon^{-}Q_{-}$ are given then by
\begin{eqnarray}
\delta\varphi&=& \varepsilon^{+}\psi_{+}+\varepsilon^{-}\psi_{-}, \nonumber
\\
\delta\psi_{+}&=&\varepsilon^{-}F, \nonumber \\
\delta\psi_{-}&=&-\varepsilon^{+}F, \nonumber \\
\delta F &=& 0,
\label{susy1}
\end{eqnarray}
while for the components of the antichiral multiplet one finds
\begin{eqnarray}
\delta\overline{\varphi}&=& 0, \nonumber
\\
\delta\overline{\psi}_{+}&=&-2i\varepsilon^{+}\partial_{+}
\overline{\varphi}, \nonumber \\
\delta\overline{\psi}_{-}&=&-2i\varepsilon^{-}\partial_{-}
\overline{\varphi}, \nonumber \\
\delta \overline{F} &=& 2i\varepsilon^{+}\partial_{+}\overline{\psi}_{-}
-2i\varepsilon^{-}\partial_{-}\overline{\psi}_{+}.
\label{susy2}
\end{eqnarray}

\section{Two-dimensional K\"ahler sigma-models}

In \cite{ck,c} an explicit construction of the nonanticommutative
deformation of two-dimensional supersymmetric theories was given. 
The resulting Lagrangian can be written as an infinite series in
$(\det{C})^{1\over 2}F^{i}$, where $F^{i}$ is the highest component
of the chiral superfield. In the following we will show how this series can be
resummed giving rise to expressions making the physical interpretation
more accessible.

The subject of our study here is the nonanticommutative deformation
(\ref{def}) of the (2,2) K\"ahler sigma-model in Euclidean space
described by the Lagrangian
\begin{eqnarray}
\mathcal{L}=\int d^{2}\theta\,d^{2}\overline{\theta}\,K(\Phi^{i},
\overline{\Phi}{}^{\,\overline{\imath}})+\int d^{2}\theta\,
\mathcal{W}(\Phi^{i})
+\int
d^{2}\overline{\theta}\,\mathcal{W}(\overline{\Phi}{}^{\,\overline{\imath}}),
\end{eqnarray}
where $\Phi^{i}$, $\overline{\Phi}{}^{\,\overline{\imath}}$
($i,\overline{\imath}=1,\ldots,N$) are respectively a set of chiral
and antichiral superfields.  We remind the reader once more that,
since we work in Euclidean space, overlines should not be interpreted
as complex conjugation. In order to construct the deformed Lagrangian
we use the prescription given in Sec. \ref{weyl} to define functions
of the superfields. 

\subsection{The superpotential}

We start with the holomorphic superpotential. Beginning with 
the following function of the Weyl symbol $\widehat{\Phi}$
\begin{eqnarray}
\mathcal{W}(\widehat{\Phi})=\sum_{n=0}^{\infty}{1\over n!}
\partial_{i_{1}}\ldots\partial_{i_{n}}\mathcal{W}(0)
\widehat{\Phi}^{i_{1}}\ldots\widehat{\Phi}^{i_{n}},
\label{series}
\end{eqnarray}
and applying the inverse Weyl map we can compute the deformed
superpotential. Therefore our first task is to compute the monomial
\begin{eqnarray}
\widehat{\Phi}^{i_{1}}\ldots\widehat{\Phi}^{i_{n}}
&=&(-1)^{n}\int d^{2}\eta_{1}\ldots\int
d^{2}\eta_{n}\,
\widetilde{\Phi}^{i_{1}}(\eta_{1})\ldots \widetilde{\Phi}^{i_{n}}(\eta_{n})
\exp\left[{-(\sum_{k=1}^{n}\eta_{k,a})\widehat{Q}^{a}}\right]\nonumber \\
&\times& \exp\left[{{1\over 2}\sum_{k<j}^{n}\langle\eta_{k},\eta_{j}\rangle}
\right],
\label{monomial}
\end{eqnarray}
where we have defined
\begin{eqnarray}
\langle\eta_{i},\eta_{j}\rangle\equiv 
C^{ab}\eta_{i,a}\eta_{j,b}={1\over M}(\eta_{i,+}
\eta_{j,-}+\eta_{i,-}\eta_{j,+}).
\end{eqnarray}
Replacing $\eta_{n}$ by a the new coordinate
$\zeta_{a}=\sum_{i=1}^{n}\eta_{i,a}$, Eq. (\ref{monomial}) can be
rewritten as
\begin{eqnarray}
\widehat{\Phi}^{i_{1}}\ldots\widehat{\Phi}^{i_{n}}&=&(-1)^{n}\int d^{2}\zeta
\,e^{-\zeta_{a}\widehat{Q}^{a}}\, 
\int d^{2}\eta_{1}\ldots\int d^{2}\eta_{n-1}\,
\widetilde{\Phi}^{i_{1}}(\eta_{1})\ldots 
\widetilde{\Phi}^{i_{n}}(\zeta-\sum_{i=1}^{n-1}\eta_{i}) \nonumber \\
&\times& \exp{\left[{1\over 2}\sum_{k<j}^{n-1}\langle\eta_{k},\eta_{j}\rangle
\right]}
\,\exp{\left[-{1\over 2}\sum_{i=1}^{n-1}\langle\zeta,\eta_{i}\rangle\right]}.
\end{eqnarray}
Finally, from this expression and using (\ref{productWeyl}) we get
\begin{eqnarray}
\widetilde{(\Phi^{i_{1}}\star\ldots\star\Phi^{i_{n}})}(\zeta)&=&(-1)^{n-1}
\int d^{2}\eta_{1}\ldots\int d^{2}\eta_{n-1}\,
\widetilde{\Phi}^{i_{1}}(\eta_{1})\ldots 
\widetilde{\Phi}^{i_{n}}(\zeta-\sum_{i=1}^{n-1}\eta_{i})
 \nonumber \\
&\times& \exp\left[{{1\over 2}\sum_{k<j}^{n-1}\langle\eta_{k},\eta_{j}\rangle
}\right]
\exp\left[{-{1\over 2}\sum_{i=1}^{n-1}\langle\zeta,\eta_{i}\rangle}\right].
\end{eqnarray}

In order to compute the contribution of the holomorphic superpotential
to the Lagrangian we need to keep only the highest component of
$\Phi^{i_{1}}(\theta)\star\ldots\star\Phi^{i_{n}}(\theta)$. Using
the identity (\ref{recipe}) this is just given by
\begin{eqnarray}
\Phi^{i_{1}}(\theta)\star\ldots\star\Phi^{i_{n}}(\theta)
\Big|_{\theta^{+}\theta^{-}}
&=& \widetilde{(\Phi^{i_{1}}\star\ldots\star\Phi^{i_{n}})}(\zeta)
\Big|_{\zeta_{\pm}=0} 
\label{int-sup}   \\
&\hspace*{-6cm}=&\hspace*{-3cm}(-1)^{n-1}
\int d^{2}\eta_{1}\ldots\int d^{2}\eta_{n-1}\,
\widetilde{\Phi}^{i_{1}}(\eta_{1})\ldots 
\widetilde{\Phi}^{i_{n}}(-\sum_{i=1}^{n-1}\eta_{i})\,
\exp\left[{{1\over 2}\sum_{k<j}^{n-1}\langle\eta_{k},\eta_{j}\rangle}\right],
\nonumber 
\end{eqnarray}
where the Fourier transforms $\widetilde{\Phi}^{i}(\eta)$ are
expressed in terms of the components of the chiral superfield
$\Phi^{i}(\theta)$ by (see Appendix A)
\begin{eqnarray}
\widetilde{\Phi}^{i}(\eta)=F^{i}+\eta_{+}\psi^{i}_{-}-
\eta_{-}\psi^{i}_{+}-\eta_{+}\eta_{-}\varphi^{i}.
\end{eqnarray}

The calculation of the integral in Eq. (\ref{int-sup}) is lengthy but
straightforward.  In particular, from Eq. (\ref{series}) we see that
only the part symmetric in the indices $\{i_{1},\ldots,i_{n}\}$
contributes.  Keeping this in mind we find the result (see Appendix B
for details)
\begin{eqnarray}
\Phi^{i_{1}}(\theta)\star\ldots\star\Phi^{i_{n}}(\theta)
\Big|_{\theta^{+}\theta^{-}}
= \sum_{k=1}^{n}F^{i_{k}}{\partial\over\partial\varphi^{i_{k}}}
\int_{-{1\over 2}}^{1\over 2}d\xi\,\left(\varphi^{i_{1}}
+{\xi\over M}F^{i_{1}}\right)
\ldots\left(\varphi^{i_{n}}+{\xi\over M}F^{i_{n}}\right)\hspace*{1cm}  
\label{firstterm}\\
-\,\,\sum_{k<\ell}^{n}\left(\psi_{+}^{i_{k}}\psi_{-}^{i_{\ell}}-
\psi_{-}^{i_{k}}\psi_{+}^{i_{\ell}}\right){\partial^{2}\over
\partial\varphi^{i_{k}}
\partial\varphi^{i_{\ell}}}
\int_{-{1\over 2}}^{1\over 2}d\xi\,\left(\varphi^{i_{1}}
+{\xi\over M}F^{i_{1}}\right)
\ldots\left(\varphi^{i_{n}}+{\xi\over M}F^{i_{n}}\right). \nonumber 
\end{eqnarray}

Plugging this into the series expansion of the superpotential leads to
the surprisingly simple result
\begin{eqnarray}
\int d^{2}\theta\,\mathcal{W}(\Phi)=F^{i}\partial_{i}\mathcal{W}_{0}(\varphi,F)
-\psi^{i}_{+}\psi^{j}_{-}\partial_{i}\partial_{j}\mathcal{W}_{0}(\varphi,F),
\label{hol}
\end{eqnarray}
where we have used the notation
\begin{eqnarray}
\mathcal{W}_{0}(\varphi,F)=\int_{-{1\over 2}}^{1\over 2}d\xi\,\,
\mathcal{W}\left(\varphi^{i}+{\xi\over M}F^{i}\right).
\label{w0}
\end{eqnarray}
The effect of the deformation on the holomorphic superpotential
amounts then to an averaging of the value of $\mathcal{W}$ around
$\varphi^{i}$ on a scale set by $F^{i}/M$. Nonanticommutativity
induces then a certain fuzziness controlled by the auxiliary field.

A similar analysis can be carried out for the antiholomorphic
superpotential.  Since the anticommutation relations of the
coordinates $\overline{\theta}^{\,\pm}$ are not deformed we will
perform the Weyl map only with respect to the holomorphic coordinates
$\theta^{\pm}$. This means that the symbols
$\widehat{\bar{\Phi}}{}^{\overline{\imath}}$ associated with the
antichiral superfields are themselves (1,1) superfields with respect
to the broken supersymmetries $\overline{Q}_{\pm}$
\begin{eqnarray}
\widehat{\bar{\Phi}}{}^{\,\overline{\imath}}
=\widehat{\bar{\sigma}}{}^{\,\overline{\imath}}
+\overline{\theta}^{\,+}
\widehat{\bar{\lambda}}_{+}{}^{\!\!\overline{\imath}}
+\overline{\theta}^{\,-}\widehat{\bar{\lambda}}_{-}{}^{\!\!\overline{\imath}}
+\overline{\theta}^{\,+}\overline{\theta}^{\,-}
\widehat{\bar{K}}{}^{\,\overline{\imath}},
\end{eqnarray}
where $\widehat{\bar{\sigma}}$, $\widehat{\bar{\lambda}}_{-}$ and
$\widehat{\bar{K}}$ are the symbols associated with the corresponding
(2,0) superfields inside of Eq. (\ref{antichiral})
\begin{eqnarray}
\overline{\sigma}{}^{\,\overline{\imath}}&=&
\overline{\varphi}{}^{\,\overline{\imath}}, 
\nonumber \\
\overline{\lambda}_{\pm}{}^{\!\!\overline{\imath}}&=& 
\overline{\psi}_{\pm}{}^{\!\!\overline{\imath}}
-2i\theta^{\pm}\partial_{\pm}\overline{\varphi}{}^{\,\overline{\imath}}, \\
\overline{K}{}^{\,\overline{\imath}}&=& \overline{F}{}^{\,\overline{\imath}}+
2i\theta^{+}\partial_{+}\overline{\psi}_{-}{}^{\!\!\overline{\imath}}+
2i\theta^{-}\partial_{-}\overline{\psi}_{+}{}^{\!\!\overline{\imath}}
+4\theta^{+}\theta^{-}
\partial_{+}\partial_{-}\overline{\varphi}{}^{\,\overline{\imath}}.\nonumber 
\end{eqnarray}
Notice that since $\overline{\sigma}{}^{\,\overline{\imath}}$ is
independent of $\theta^{\pm}$ its symbol is just the c-number
$\overline{\varphi}{}^{\,\overline{\imath}}$ itself.

The antiholomorphic superpotential is given by the expansion
\begin{eqnarray}
\overline{\mathcal{W}}(\widehat{\bar{\Phi}})=\sum_{n=0}^{\infty}{1\over n!}
{\partial}_{\overline{\imath}_{1}}\ldots
{\partial}_{\overline{\imath}_{n}}\overline{\mathcal{W}}(0)
\widehat{\bar{\Phi}}{}^{\,\overline{\imath}_{1}}\ldots
\widehat{\bar{\Phi}}{}^{\,\overline{\imath}_{n}}.
\label{series2}
\end{eqnarray}
Here, however, only the part proportional to
$\overline{\theta}^{\,+}\overline{\theta}^{\,-}$ contributes to the
Lagrangian. Therefore the only relevant terms in Eq. (\ref{series2})
are the ones of the form
\begin{eqnarray}
\widehat{\bar{K}}{}^{\,\overline{\imath}_{1}} \,
\widehat{\bar{\sigma}}{}^{\,\overline{\imath}_{2}}\ldots
\widehat{\bar{\sigma}}{}^{\,\overline{\imath}_{n}},
\hspace*{1cm}
\widehat{\bar{\lambda}}_{+}{}^{\!\!\overline{\imath}_{1}}\,
\widehat{\bar{\lambda}}_{-}{}^{\!\!\overline{\imath}_{2}}\,
\widehat{\bar{\sigma}}{}^{\,\overline{\imath}_{3}}\ldots
\widehat{\bar{\sigma}}{}^{\,\overline{\imath}_{n}}
\end{eqnarray}
and permutations. We can proceed then as in the holomorphic case by
identifying the corresponding Fourier transforms.  Unlike the
calculation of the holomorphic part, here we are interested in the
lowest component in $\theta^{+}\theta^{-}$. From
Eq. (\ref{transform-comp}) we see that the relevant terms are the
symmetric parts of
\begin{eqnarray*}
-\widetilde{\left({\bar{K}}{}^{\,\overline{\imath}_{1}}\star 
{\bar{\sigma}}{}^{\,\overline{\imath}_{2}}\star\ldots\star
{\bar{\sigma}}{}^{\,\overline{\imath}_{n}}\right)}(\zeta)
\Big|_{\zeta_{+}\zeta_{-}},
\hspace*{1cm}
-\widetilde{\left({\bar{\lambda}}_{+}{}^{\!\!\overline{\imath}_{1}}\star
{\bar{\lambda}}_{-}{}^{\!\!\overline{\imath}_{2}}\star
{\bar{\sigma}}^{\overline{\imath}_{3}}\star\ldots\star
{\bar{\sigma}}^{\overline{\imath}_{n}}\right)}(\zeta)
\Big|_{\zeta_{+}\zeta_{-}}.
\end{eqnarray*}
Proceedings in this way we find
\begin{eqnarray}
\overline{\Phi}{}^{\,\overline{\imath}_{1}}\star\ldots\star
\overline{\Phi}{}^{\,\overline{\imath}_{n}}
\Big|_{\theta^{+}\theta^{-},{\rm sym}}
&=& \sum_{k=1}^{n}
\overline{F}{}^{\,\overline{\imath}_{k}}
{\partial\over\partial\overline{\varphi}^{\,\overline{\imath}_{k}}}
\left(\overline{\varphi}{}^{\,\overline{\imath}_{1}}
\ldots\overline{\varphi}{}^{\,\overline{\imath}_{n}}\right) \nonumber \\
&-&\sum_{k<\ell}^{n}\left(\overline{\psi}_{+}{}^{\!\!\overline{\imath}_{k}}\,
\overline{\psi}_{-}{}^{\!\!\overline{\imath}_{\ell}}-
\overline{\psi}_{-}{}^{\!\!\overline{\imath}_{k}}\,
\overline{\psi}_{+}{}^{\!\!\overline{\imath}_{\ell}}\right)
{\partial^{2}\over\partial\overline{\varphi}{}^{\,\overline{\imath}_{k}}
\partial\overline{\varphi}{}^{\,\overline{\imath}_{\ell}}}
\left(\overline{\varphi}{}^{\,\overline{\imath}_{1}}
\ldots\overline{\varphi}{}^{\,\overline{\imath}_{n}}\right).
\end{eqnarray}
The first interesting thing is that all dependence in the deformation
scale disappears.  Actually all terms depending on $M$ disappear after
symmetrization. Therefore the antiholomorphic superpotential does not
suffer any deformation and we retrieve the standard expression
\begin{eqnarray}
\int d^{2}\overline{\theta}\,\overline{\mathcal{W}}(\overline{\Phi})=
\overline{F}{}^{\,\overline{\imath}}\,{\partial}_{\overline{\imath}}
\,\overline{\mathcal{W}}(\overline{\varphi})
-\overline{\psi}{}^{\,\overline{\imath}}_{+}\,
\overline{\psi}{}^{\,\overline{\jmath}}_{-}\,
{\partial}_{\overline{\imath}}
\,{\partial}_{\overline{\jmath}}\,
\overline{\mathcal{W}}(\overline{\varphi}).
\label{antihol}
\end{eqnarray}

Equations (\ref{hol}) and (\ref{antihol}) show that the noncommutative
deformation only affects the holomorphic part of the superpotential
\cite{seiberg}. It is quite remarkable, however, that the particular
deformation suffered by the holomorphic superpotential has a clear
geometric interpretation as smearing in the target space coordinates.

\subsection{The K\"ahler potential}

After the analysis of the superpotential we turn our attention to the
deformation of the K\"ahler potential. Using Eqs. (\ref{Q}) and 
(\ref{productWeyl}) we construct the corresponding K\"ahler function 
$K(\widehat{\Phi},\widehat{\bar{\Phi}})$ for the symbols
$\widehat{\Phi}^{i}$, $\widehat{\bar{\Phi}}{}^{\overline{\jmath}}$ as
\begin{eqnarray}
K(\widehat{\Phi},\widehat{\bar{\Phi}})=
\sum_{n,m=0}^{\infty}{1\over (n+m)!}\partial_{i_{1}}\ldots\partial_{i_{n}}
{\partial}_{\overline{\jmath}_{1}}\ldots
{\partial}_{\overline{\jmath}_{m}}
K(0,0)\,\widehat{\Phi}^{(i_{1}}\ldots\widehat{\Phi}^{i_{n}}
\widehat{\bar{\Phi}}{}^{\,\overline{\jmath}_{1}}\ldots
\widehat{\bar{\Phi}}{}^{\,\overline{\jmath}_{m})},
\label{kahler1}
\end{eqnarray}
where we have to consider all the possible permutations of
$\widehat{\Phi}^{i}$'s and
$\widehat{\bar{\Phi}}{}^{\overline{\imath}}$'s. As in the case of our
discussion of the antiholomorphic superpotential, the fact that only
$\overline{\theta}^{\,+}\overline{\theta}^{\,-}$ contributes to the
Lagrangian implies that in computing the different monomials in
Eq. (\ref{kahler1}) the only terms that we have to take into consideration 
are the ones of the form
\begin{eqnarray}
\widehat{\Phi}^{i_{1}}\ldots \widehat{\Phi}^{i_{n}}
\widehat{\bar{K}}{}^{\,\overline{\imath}_{1}} \,
\widehat{\bar{\sigma}}{}^{\,\overline{\imath}_{2}}\ldots
\widehat{\bar{\sigma}}{}^{\,\overline{\imath}_{m}},
\hspace*{1cm}
\widehat{\Phi}^{i_{1}}\ldots \widehat{\Phi}^{i_{n}}
\widehat{\bar{\lambda}}_{+}{}^{\!\!\overline{\imath}_{1}}\,
\widehat{\bar{\lambda}}_{-}{}^{\!\!\overline{\imath}_{2}}\,
\widehat{\bar{\sigma}}{}^{\,\overline{\imath}_{3}}\ldots
\widehat{\bar{\sigma}}{}^{\,\overline{\imath}_{m}}
\end{eqnarray}
in all possible orderings. The calculation of the Lagrangian can now
be carried out using the same techniques used to evaluate the superpotential
(see Appendix B). First the Fourier transform of the corresponding
monomials is evaluated. The term contributing to the Lagrangian
corresponds then to the lowest component of the Fourier transform,
according to Eq. (\ref{recipe}). At the end, the resulting Lagrangian
can be written
\begin{eqnarray}
\int d^{2}\theta\,d^{2}\overline{\theta}
\,K(\Phi,\overline{\Phi})=\mathcal{L}_{0}+\mathcal{L}_{1},
\label{L}
\end{eqnarray}
where 
\begin{eqnarray}
\mathcal{L}_{0}&=& 4\,{\partial}_{\overline{\jmath}}\,
K_{0}(\varphi,F,\overline{\varphi})\,
\partial_{+}\partial_{-}\overline{\varphi}^{\,\overline{\jmath}}+
\partial_{i}\,{\partial}_{\overline{\jmath}}\,
K_{0}(\varphi,F,\overline{\varphi})\,
\left(2i\psi_{+}^{i}\,\partial_{-}\,
\overline{\psi}_{-}{}^{\!\!\overline{\jmath}}
+2i\psi_{-}^{i}\,\partial_{+}\,
\overline{\psi}_{-}{}^{\!\!\overline{\jmath}}+F^{i}\,
\overline{F}^{\,\overline{\jmath}}\right) \nonumber \\
&+&4\,{\partial}_{\overline{\jmath}}\,
{\partial}_{\overline{\ell}}\,
K_{0}(\varphi,F,\overline{\varphi})\,
\partial_{+}\overline{\varphi}{}^{\,\overline{\jmath}}\,
\partial_{-}\,\overline{\varphi}{}^{\,\overline{\ell}}
-\partial_{i}\,\partial_{k}\,
{\partial}_{\overline{\jmath}}\,K_{0}(\varphi,F,\overline{\varphi})\,
\psi_{+}^{i}\,\psi_{-}^{k}\,\overline{F}{}^{\,\overline{\jmath}} \nonumber \\
&+& \partial_{i}\,{\partial}_{\overline{\jmath}}\,
{\partial}_{\overline{\ell}}\,
K_{0}(\varphi,F,\overline{\varphi})\,
\left(2i\,\psi_{+}^{i}\,\overline{\psi}_{+}{}^{\!\!\overline{\jmath}}\,
\partial_{-}\overline{\varphi}{}^{\overline{\ell}}+
2i\,\psi_{-}^{i}\,\overline{\psi}_{-}{}^{\!\!\overline{\jmath}}\,
\partial_{+}\overline{\varphi}{}^{\overline{\ell}}
-F^{i}\,\overline{\psi}_{+}{}^{\!\!\overline{\jmath}}\,
\overline{\psi}_{-}{}^{\!\!\overline{\ell}} \right) \nonumber \\
&+&\partial_{i}\,\partial_{k}\,{\partial}_{\overline{\jmath}}\,
{\partial}_{\overline{\ell}}\,K_{0}(\varphi,F,\overline{\varphi})\,
\psi_{+}^{i}\,\psi_{-}^{k}\,
\overline{\psi}_{+}{}^{\overline{\!\!\overline{\jmath}}}\,
\overline{\psi}_{-}{}^{\overline{\!\!\overline{\ell}}},
\label{L1}
\end{eqnarray}
and
\begin{eqnarray}
\mathcal{L}_{1}&=&{4\over M}\partial_{i}\,
{\partial}_{\overline{\jmath}}
\,K_{1}(\varphi,F,\overline{\varphi})\,F^{i}\,\partial_{+}\partial_{-}
\overline{\varphi}^{\overline{\jmath}}-
{4\over M}\partial_{i}\partial_{k}\,{\partial}_{\overline{\jmath}}
\,K_{1}(\varphi,F,\overline{\varphi})\,\psi_{+}^{i}\,\psi_{-}^{k}
\partial_{+}\partial_{-}\overline{\varphi}^{\,\overline{\jmath}}  
\label{L2}\\
&+&{4\over M}\partial_{i}\,{\partial}_{\overline{\jmath}}
\,{\partial}_{\overline{\ell}}
K_{1}(\varphi,F,\overline{\varphi})\,F^{i}\,
\partial_{+}\overline{\varphi}^{\,\overline{\jmath}}
\partial_{-}\overline{\varphi}^{\,\overline{\ell}}
-{4\over M}\partial_{i}\,\partial_{k}\,
{\partial}_{\overline{\jmath}}\,
{\partial}_{\overline{\ell}}\,K_{1}(\varphi,F,\overline{\varphi})\,
\psi_{+}^{i}\,\psi_{-}^{k}\,
\partial_{+}\overline{\varphi}^{\,\overline{\jmath}}
\partial_{-}\overline{\varphi}^{\,\overline{\ell}}.
\nonumber 
\end{eqnarray}
Here we have used a similar notation to the one used in the expression
of the superpotential and defined
\begin{eqnarray}
K_{m}(\varphi,F,\overline{\varphi})=
\int_{-{1\over 2}}^{1\over 2}\xi^{m}\,d\xi\,
K\left(\varphi+{\xi\over M}F,\overline{\varphi}
\right).
\end{eqnarray}
Here, as in the case of the holomorphic superpotential (\ref{hol}),
$\partial_{i}$, ${\partial}_{\overline{\jmath}}$ are
understood as derivative with respect to $\varphi^{i}$ and 
$\overline{\varphi}^{\overline{\jmath}}$ respectively.

Thus, we have found that the infinite series in \cite{c} for the
K\"ahler potential of the nonanticommutative (2,2) sigma-model can be
nicely resummed. Actually, we find that the resulting Lagrangian can
be divided in two parts. The first one, $\mathcal{L}_{0}$ is
identical, up to total derivatives, to the standard (2,2) Lagrangian
\cite{zumino} with the only replacement of the K\"ahler potential
$K(\varphi,\overline{\varphi})$ by the smeared function
$K_{0}(\varphi,F,\overline{\varphi})$. This is exactly the same
smearing found in the holomorphic superpotential.  On the other hand
$\mathcal{L}_{1}$ contains a number of dimension 3 operators
multiplying what one might call the first moment of the smeared
K\"ahler potential, $K_{1}(\varphi,F,\overline{\varphi})$.

One can easily check that the Lagrangian is invariant under the
supersymmetry transformations (\ref{susy1})-(\ref{susy2}) generated by
$Q_{\pm}$. Remarkably, $\mathcal{L}_{0}$ and $\mathcal{L}_{1}$ are
{\em independently} invariant under the residual
supersymmetry\footnote{Actually, the first and second term in
Eq. (\ref{L1}) are together invariant under $Q_{\pm}$, as well as the
third and fourth terms combined.}. Therefore the $Q_{\pm}$
supersymmetries do not act irreducibly on the Lagrangian obtained
using the standard prescription to implement the nonanticommutative
deformation. On the other hand under a K\"ahler transformation
$K\rightarrow K+f(\varphi)+\overline{f}(\overline{\varphi})$ the
Lagrangian transforms by a total derivative
\begin{eqnarray}
\mathcal{L} \longrightarrow \mathcal{L}+4\partial_{+}
\left[\,{\partial}_{\overline{\jmath}}\,\,
\overline{f}(\overline{\varphi})
\,\partial_{-}\overline{\varphi}{}^{\,\overline{\jmath}}\,\right].
\end{eqnarray}
Notice that this total derivative only involves antiholomorphic fields
and therefore is insensitive to the deformation.

We should point out that the full Lagrangian
$\mathcal{L}_{0}+\mathcal{L}_{1}$ can be written solely in terms of
the function $K_{0}(\varphi,F,
\overline{\varphi})$ due to the identity
\begin{eqnarray}
{1\over M}\partial_{i}K_{m}(\varphi,F,\overline{\varphi})=
{\partial\over\partial F^{i}}K_{m-1}(\varphi,F,\overline{\varphi}),
\end{eqnarray}
at the price, however, of introducing derivatives of the deformed
K\"ahler potential with respect to the auxiliary field. Since all
dependence of $K_{0}(\varphi,F,\overline{\varphi})$ on $F^{i}$
disappears when $M\rightarrow\infty$, only $\mathcal{L}_{0}$ survives
in the anticommutative limit, with $K_{0}(\varphi,F,
\overline{\varphi})$ replaced by the undeformed K\"ahler potential
$K(\varphi,\overline{\varphi})$.
In this way the standard K\"ahler sigma model is retrieved \cite{zumino}.

As advertised in the Introduction, the na\"{\i}ve generalization
of Zumino's Lagrangian (\ref{L1}) is invariant with respect to the
residual supersymmetry and also compatible with the deformed superspace
structure. This robustness was not expected from the start and it is
surprising that after all the dust settles we end up with a well-known
Lagrangian. However, a number of questions arise. A first one is
how the usual Ricci-flat condition for the vanishing of the beta-function
\cite{beta} changes. Also, the structure of the sigma-model instantons
is likely to be modified, for example in the case of the 
$\mathbb{CP}^{1}$ model.

\section{The classical structure of vacua}

In the undeformed case, the analysis of the (traslationally invariant)
vacuum structure of the theory begins with the study of the effective
potential for the scalar fields $V(\phi)_{\rm eff}$. Its critical
points describe then the possible vacua of the theory. Hermiticity of
the original theory guarantees that the potential obtained is positive
definite, so that the vanishing of $V(\phi)_{\rm eff}$ implies the 
existence of a supersymmetric vacuum. In the cases of interest for our
analysis the order parameter is the vacuum expectation value of the
auxiliary field $F^{i}$. 

However, once hermiticity is lost it is difficult to obtain general
properties. Let us illustrate the point with a few examples. In the first
one we start with a generic superpotential $\mathcal{W}(\Phi)$ and a rather
simple K\"ahler potential
\begin{eqnarray}
K(\Phi,\overline{\Phi})=\delta_{i\overline{\jmath}}\Phi^{i}\overline{\Phi}
{}^{\,\overline{\jmath}}.
\end{eqnarray}
This K\"ahler potential does not receive any deformation so the
kinetic terms of all component fields of $\Phi^{i}$ are the standard
ones. Hence, the only bosonic terms without space-time derivatives in
the Lagrangian are
\begin{eqnarray}
\delta_{i\overline{\jmath}}F^{i}\overline{F}{}^{\,\overline{\jmath}}
+F^{i}\partial_{i}\mathcal{W}_{0}(\varphi,F)+
\overline{F}{}^{\,\overline{\jmath}}
{\partial}_{\overline{\jmath}}\overline{\mathcal{W}}(\overline{\varphi}),
\label{veff}
\end{eqnarray}
with $\mathcal{W}_{0}(\varphi,F)$ given by Eq. (\ref{w0}). Varying with
respect to $\overline{F}{}^{\,\overline{\imath}}$ we obtain 
\begin{eqnarray}
F^{i}+\delta^{i\overline{\jmath}}\,{\partial}_{\overline{\jmath}}
\overline{\mathcal{W}}(\overline{\varphi})=0,
\label{F}
\end{eqnarray}
and substituting  $F^{i}$ in Eq. (\ref{veff}), as given by the last equation, 
we find (cf. \cite{HKKS})
\begin{eqnarray}
V(\varphi,\overline{\varphi})_{\rm eff}=
{\partial}_{\overline{\jmath}}\,\overline{\mathcal{W}}(\overline{\varphi})\,
\delta^{i\overline{\jmath}}\,\partial_{i}\,\mathcal{W}_{0}
\left(\varphi,-{1\over M}{\overline{\partial}}\,
\overline{\mathcal{W}}\right).
\label{defW}
\end{eqnarray}
It is straightforward to check that in the anticommutative limit $M
\rightarrow\infty$ the standard result is recovered.

It is clear from (\ref{defW}) that if there are values of
$\langle\overline{\varphi}{}^{\,\overline{\jmath}}\rangle$ solving
Eq. (\ref{F}) for which $\langle F^{i}\rangle=0$ then Eq. (\ref{defW})
receives no deformation and the vacuum of the theory will be the same
as for the undeformed theory.  However, the effective potential
(\ref{defW}) for the scalars is not positive definite.  In fact it is
complex and its real part does not seem to have any positivity
property in general. The analysis of the behavior of the theory at the
critical point should be carried out using a saddle point
analysis. Since we are working in Euclidean space, it is not clear
what this means for the realization of $\mathcal{N}=1/2$
supersymmetry.

The second example we want to consider here involves a non-trivial
K\"ahler potential which receives a deformation. We have seen that the
full Lagrangian obtained using the standard prescription to implement
the nonanticommutative deformation is not irreducible, and that a
simple deformation of Zumino's Lagrangian \cite{zumino} as in
Eq. (\ref{L1}) is also invariant under the $\mathcal{N}=1/2$
supersymmetry transformations (\ref{susy1}) and (\ref{susy2}). If we
are interested in translationally invariant ground states (or critical
points) the additional terms (\ref{L2}) do not play any r\^{o}le. In
this case the changes on the previous analysis to include a
non-trivial K\"ahler potential are straightforward. If we drop in the
Lagrangian all derivative terms\footnote{We have not analyzed here
``stripped'' states \cite{stripped} although they are very likely to
appear in this context. In the following paragraphs we consider only
space-time independent vacuum expectation values.} we obtain
\begin{eqnarray}
\mathcal{L}=G_{i\overline{\jmath}}(\varphi,F,\overline{\varphi})
F^{i}\overline{F}{}^{\,\overline{\jmath}}+F^{i}{\partial}_{i}
\mathcal{W}_{0}(\varphi,F)+\overline{F}{}^{\,\overline{\jmath}}\,
{\partial}_{\overline{\jmath}}
\overline{\mathcal{W}}(\overline{\varphi}),
\end{eqnarray}
where, according to (\ref{L1}),
\begin{eqnarray}
G_{i\overline{\jmath}}(\varphi,F,\overline{\varphi})=
\partial_{i}{\partial}_{\overline{\jmath}}
K_{0}(\varphi,F,\overline{\varphi})
\end{eqnarray}
is the deformed K\"ahler metric. Varying with respect to $\overline{F}
{}^{\,\overline{\jmath}}$ we obtain
\begin{eqnarray}
G_{i\overline{\mathcal{\jmath}}}(\varphi,F,\overline{\varphi})
F^{i}+{\partial}_{\overline{\jmath}}
\overline{\mathcal{W}}(\overline{\varphi})=0.
\label{FG}
\end{eqnarray}
In principle, it would be possible to solve this equation for 
$F^{i}$ and substituting in the Lagrangian we would obtain 
the effective scalar potential
\begin{eqnarray}
V(\varphi,\overline{\varphi})_{\rm eff}=G^{i\overline{\jmath}}\Big(\varphi,
F(\varphi,\overline{\varphi}),\overline{\varphi}\Big)\,
\partial_{i}\mathcal{W}_{0}\Big(\varphi,F(\varphi,\overline{\varphi})\Big)
{\partial}_{\overline{\jmath}}
\overline{\mathcal{W}}(\overline{\varphi}),
\label{VG}
\end{eqnarray}
with $F^{i}(\varphi)$ solving Eq. (\ref{FG}). However, this is not a very
illuminating expression, since getting solving for $F^{i}$ in 
(\ref{FG}) is in general complicated. Once again (\ref{VG}) is 
complex. It would be interesting to analyze in more detail the behavior
of the scalar theory close to its critical points in some examples.
We will come back to this issue in the future.

\section{Concluding remarks}

We have seen that nonanticommutative two-dimensional sigma models
admit a closed form in which the deformation affects the K\"ahler
potential and the superpotential. Physically, this deformation
corresponds to a smearing of the target space holomorphic coordinates.
According to this, the holomorphic superpotential is obtained by
averaging its undeformed value between $\varphi^{i}-F^{i}/(2M)$ and
$\varphi^{i}+F^{i}/(2M)$ as shown in Eq. (\ref{w0}). It is important
to stress that although we have derived this relation in two
dimensions a similar expression would hold also for the superpotential
of the four-dimensional $\mathcal{N}=1/2$ Wess-Zumino model. The
expressions for the scalar potential found in Refs. \cite{BFR,HKKS}
can be actually retrieved using the identity
\begin{eqnarray}
{F^{i}\over M}\partial_{i}\mathcal{W}_{0}(\varphi,F)=
\mathcal{W}\left(\varphi+{F\over 2M}\right)
-\mathcal{W}\left(\varphi-{F\over 2M}\right).
\end{eqnarray}

In the case of the deformation of the kinetic part of the sigma-model
action we found that it consists of two parts. The first one is just
the usual K\"ahler action form with the K\"ahler potential deformed
to\footnote{We consider the undeformed Zumino's Lagrangian with a
kinetic term for the scalar fields of the form
$\partial_{\overline{\imath}}K(\varphi,\overline{\varphi})
\partial_{+}\partial_{-}
\overline{\varphi}{}^{\overline{\imath}}$, which differs from the
Lagrangian appearing in Ref. \cite{zumino} by a total derivative.}
\begin{eqnarray}
K_{0}(\varphi,F,\overline{\varphi})=\int_{-{1\over 2}}^{1\over 2}d\xi
\,K\left(\varphi+{\xi\over M}F,\overline{\varphi}\right).
\end{eqnarray}
Together with this, there is a second term which contains higher 
dimensional operators suppressed by $1/M$ and with couplings
governed by the function
\begin{eqnarray}
K_{1}(\varphi,F,\overline{\varphi})=\int_{-{1\over 2}}^{1\over 2}\xi d\xi
\,K\left(\varphi+{\xi\over M}F,\overline{\varphi}\right).
\end{eqnarray}  

It is quite remarkable that our construction allows us to write the action
as a term which corresponds to the usual (2,2) action with
a deformed K\"ahler potential $K_{0}(\varphi,F,\overline{\varphi})$
together a few higher dimensional terms. When written in the
form (\ref{L1}) it is obvious that $F^{i}$ remains a non-propagating field,
in spite of its now more complicated couplings. Of course the action
$\mathcal{L}_{0}+\mathcal{L}_{1}$ can be written, modulo total derivatives,
in the canonical form with a kinetic term for the scalars of the form
$\partial_{i}{\partial}_{\overline{\jmath}}K_{0}(\varphi,F,
\overline{\varphi})(\partial_{+}\varphi^{i}\partial_{-}
\overline{\varphi}{}^{\,\overline{\jmath}}+
\partial_{-}\varphi^{i}\partial_{+} 
\overline{\varphi}{}^{\,\overline{\jmath}})$. However, due to the
extra dependence of $K_{0}(\varphi,F,\overline{\varphi})$
there will be new terms in $\mathcal{L}_{1}$
containing one derivative of the auxiliary field $\partial_{\pm}F^{i}$.

As we pointed out, the (2,2) nonanticommutative sigma model can be obtained
from the corresponding $\mathcal{N}=1/2$ four-dimensional theory. Then
from the analysis of Ref. \cite{seiberg} it follows that an antichiral
ring is preserved. 

Hence, in this article we have shown that the $\mathcal{N}=2$
nonanticommutative sigma-models, in spite of the algebraic
complications, can be written as a generalization of Zumino's
Lagrangian \cite{zumino}. A number of issues, however, remain to be
addressed. A first one concerns the quantum structure of the theory
and, in particular, the conditions for the vanishing of the
beta-function at one-loop. In the undeformed case the beta-function
vanish at one loop provided the target space manifold is Ricci
flat \cite{beta}. It will be interesting to understand how this condition is
changed in the nonanticommutative case where, as we have seen above,
the two-dimensional deformation induces a fuzziness in the target
manifold.

Since nonanticommutative theories are naturally defined in Euclidean
space, a second question that can be addressed is about the analog of
two-dimensional instantons in the deformed sigma model. Actually, concerning
the Euclidean character of the theory it would be interesting to see 
whether there is any way to overcome the constraints of Ref. \cite{low}
to define models in Lorentzian space-time (see \cite{lorentzian} for 
some analysis in this direction). This is specially interesting 
in order to see if the centrally extended superalgebra (\ref{p2}) 
can play any r\^{o}le in Lorentzian Quantum Field Theory.

\section*{Acknowledgments}

We thank Jos\'e Barb\'on, Costas Kounnas, Kerstin Kunze and 
Juan Ma\~nes for useful discussions. M.A.V.-M. thanks 
CERN Theory Division for hospitality. The work of M.A.V.-M. has been
partially supported by Spanish Science Ministry Grants  FPA 2002-02037 and 
BFM2003-02121. 

\section*{Appendix A. Fourier transforms with Grassmann variables}
\renewcommand{\thesection}{A}

In this Appendix we summarize some basic results of the theory of
Fourier transforms for anticommuting variables. Given a function
$f(\theta)$ depending on $2n$ anticommuting variables $\theta^{1},\ldots
\theta^{2n}$, its Fourier transform is defined as
\begin{eqnarray}
\widetilde{f}(\eta)=\int d^{2n}\theta\,e^{\eta_{a}\theta^{a}}f(\theta),
\label{fourier}
\end{eqnarray}
where the phase of the integration measure is fixed by requiring 
\begin{eqnarray}
\int d^{2n}\theta\,\, \theta^{1}\ldots\theta^{2n}=1.
\end{eqnarray}
Using that $\delta^{(2n)}(\theta)=\theta^{1}\ldots\theta^{2n}$ it is 
easily proved that the inverse Fourier transform is given by
\begin{eqnarray}
f(\theta)=(-1)^{n}\int d^{2n}\eta\,e^{-\eta_{a}\theta^{a}}\,
\widetilde{f}(\eta),
\end{eqnarray}
and the delta function can be represented as
\begin{eqnarray}
\delta^{(2n)}(\theta)=(-1)^{n}\int d^{2n}\eta\,e^{\eta_{a}\theta^{a}}.
\end{eqnarray}

Let us analyze now the case of two-dimensional superspace with
coordinates $y^{\pm}$, $\theta^{\pm}$. The definition of the Fourier
transform simplifies to
\begin{eqnarray}
\widetilde{f}(\eta_{\pm})=\int d^{2}\theta\,\,
e^{\eta_{+}\theta^{+}+\eta_{-}\theta^{-}}
f(\theta^{\pm}),
\end{eqnarray}
while the inversion formula is 
\begin{eqnarray}
f(\theta)=-\int d^{2}\eta\,\,e^{-(\eta_{+}\theta^{+}+\eta_{-}\theta^{-})}
\widetilde{f}(\eta_{\pm}).
\end{eqnarray}
Given a general function in superspace it can be decomposed as
\begin{eqnarray}
f(y^{\pm},\theta^{\pm})=f_{0}(y^{\pm})+\theta^{+}f_{+}(y^{\pm})
+\theta^{-}f_{-}(y^{\pm})+\theta^{+}\theta^{-}f_{1}(y^{\pm}).
\end{eqnarray}
By applying now the definition (\ref{fourier}) one finds that its
Fourier transform $\widetilde{f}(y^{\pm},\eta_{\pm})$ with respect to
the anticommuting coordinates is given by
\begin{eqnarray}
\widetilde{f}(y^{\pm},\eta_{\pm})=f_{1}(y^{\pm})+\eta_{+}f_{-}(y^{\pm})
-\eta_{-}f_{+}(y^{\pm})-\eta_{+}\eta_{-}f_{0}(y^{\pm}).
\label{transform-comp}
\end{eqnarray}
That is, upon Fourier transformation with respect to the fermionic
coordinates the components of a (2,2) superfield reshuffle. In
particular, the highest and lowest components interchange, which
implies the useful identities
\begin{eqnarray}
f(y^{\pm},\theta^{\pm})\Big|_{\theta^{+}\theta^{-}}&=&
\widetilde{f}(y^{\pm},\eta_{\pm})\Big|_{\eta_{\pm}=0}, \\ \nonumber
f(y^{\pm},\theta^{\pm})\Big|_{\theta^{\pm}=0}&=&
-\widetilde{f}(y^{\pm},\eta_{\pm})\Big|_{\eta_{+}\eta_{-}}.
\label{recipe}
\end{eqnarray}

\section*{Appendix B. Some details of the calculation of the deformed 
Lagrangian}
\renewcommand{\thesection}{B}
\setcounter{equation}{0}

Here we outline part of the calculations of Section 4 leading to the
Lagrangian of the deformed (2,2) sigma-model. In order to compute the
relevant fermionic integrals we are going to forget about the target
space indices and use the simplified notation
\begin{eqnarray}
\varphi^{(k)}\equiv \varphi^{i_{k}}, \hspace*{1cm}
\psi_{\pm}^{(k)} \equiv \psi_{\pm}^{i_{k}}, 
\hspace*{1cm} F^{(k)}\equiv F^{i_{k}}.
\end{eqnarray}
Therefore the Fourier transform of the chiral superfield $\tilde{\Phi}^{i_{k}}$
can be written as
\begin{eqnarray}
\widetilde{\Phi}^{i_{k}}\equiv  \widetilde{\Phi}^{(k)}&=&F^{(k)}
+\eta_{+}\psi_{-}^{(k)}-\eta_{-}\psi_{+}^{(k)}-\eta_{+}\eta_{-}
\varphi^{(k)} \nonumber \\
&=& F^{(k)}\left[1-\eta_{+}\eta_{-}{\varphi^{(k)}\over F^{(k)}}
+\eta_{+}{\psi_{-}^{(k)}\over F^{(k)}}-\eta_{-}{\psi_{+}^{(k)}\over F^{(k)}}
\right] \nonumber \\
&\equiv & F^{(k)}\left[1-\eta_{+}\eta_{-}{\hat{\varphi}^{(k)}}
+\eta_{+}{\hat{\psi}_{-}^{(k)}}-\eta_{-}{\hat{\psi}_{+}^{(k)}}
\right],
\label{a1}
\end{eqnarray}
where we have used the notation 
\begin{eqnarray}
\hat{\varphi}^{(k)}\equiv {\varphi^{(k)}\over F^{(k)}},
\hspace*{1cm} \hat{\psi}_{\pm}^{(k)}\equiv {\psi_{\pm}^{(k)}\over F^{(k)}}.
\end{eqnarray}
It is important not to confuse this notation with the one for the Weyl
symbols introduced in Section 2. Now Eq. (\ref{a1}) can be written as
\begin{eqnarray}
\widetilde{\Phi}^{(k)}=F^{(k)}e^{-\eta_{+}\eta_{-}\hat{\varphi}^{(k)}}
\left[1+\eta_{+}\hat{\psi}_{-}^{(k)}-\eta_{-}\hat{\psi}_{+}^{(k)}\right].
\label{hats}
\end{eqnarray}
Hence, the integral in Eq. (\ref{int-sup}) can be expressed as
\begin{eqnarray}
& & 
(-1)^{n-1}F^{(1)}\ldots F^{(n)}\Big[\int d^{2}\eta_{1}\ldots 
\int d^{2}\eta_{n-1}
e^{\boldsymbol{\eta}_{+}^{T}\,\mathcal{D}\,\boldsymbol{\eta}_{-}} 
\nonumber \\
& & \hspace*{5cm} 
+ \sum_{i<j}^{n}\int d^{2}\eta_{1}\ldots \int d^{2}\eta_{n-1}
\Psi^{(i)}\Psi^{(j)}\,e^{\boldsymbol{\eta}_{+}^{T}\,\mathcal{D}\,
\boldsymbol{\eta}_{-}}\Big].
\label{integrals}
\end{eqnarray}
Here we have used the vector notation $(\boldsymbol{\eta}_{\pm})_{i}
=\eta_{i,\pm}$, 
\begin{eqnarray}
\Psi^{(k)}&=& \eta_{k,+}\psi_{+}^{(k)}-\eta_{k,-}\psi_{-}^{(k)}, \hspace*{5cm}
k=1,\ldots,n-1, \nonumber \\
\Psi^{(n)}&=& \left(-\sum_{i=1}^{n-1}\eta_{i,+}\right)\psi_{+}^{(n)}
-\left(-\sum_{i=1}^{n-1}\eta_{i,-}\right)\psi_{-}^{(n)},
\end{eqnarray}
and $\mathcal{D}$ is a $(n-1)\times (n-1)$ matrix given by
\begin{eqnarray}
\mathcal{D}=\left(
\begin{array}{cccc}
-\hat{\varphi}^{(1)}-\hat{\varphi}^{(n)} & {1\over 2M}-\hat{\varphi}^{(n)}
&\cdots  & {1\over 2M}-\hat{\varphi}^{(n)} \\
 & & &  \\
-{1\over 2M}-\hat{\varphi}^{(n)} & -\hat{\varphi}^{(2)}-\hat{\varphi}^{(n)} 
&\cdots   & {1\over 2M}-\hat{\varphi}^{(n)}
\\ & & & \\
\vdots  & \vdots & &\vdots \\
& &  & \\
-{1\over 2M}-\hat{\varphi}^{(n)} & -{1\over 2M}-\hat{\varphi}^{(n)} &\cdots  & 
-\hat{\varphi}^{(n-1)}-\hat{\varphi}^{(n)}
\end{array}\right).
\label{matrix}
\end{eqnarray}

The first integral within the brackets in Eq. (\ref{integrals}) is equal to
\begin{eqnarray}
\det{\,\mathcal{D}}&=&(-1)^{n-1}\int_{-{1\over 2}}^{1\over 2}
d\xi\,P_{n-1}\left(\hat{\varphi}^{(1)}+{\xi\over M},\ldots,
\hat{\varphi}^{(n)}+{\xi\over M}\right) \nonumber \\
&=& (-1)^{n-1}\sum_{i=1}^{n}{\partial\over\partial\hat{\varphi}^{(i)}}
\int_{-{1\over 2}}^{1\over 2}
d\xi\,\left(\hat{\varphi}^{(1)}+{\xi\over M}\right)\ldots
\left(\hat{\varphi}^{(n)}+{\xi\over M}\right),
\label{det}
\end{eqnarray}
where 
\begin{eqnarray}
P_{k}(x_{1},\ldots,x_{n})=\sum_{i_{1}<\ldots<i_{k}}^{n}
x_{i_{1}}\ldots x_{i_{k}}
\end{eqnarray}
are the elementary symmetric polynomials of degree $k$. The
computation of the second integral in Eq. (\ref{integrals}), that we
denote by $\langle \Psi^{(i)}\Psi^{(j)}\rangle$, is a bit more
involved. In particular we have to keep in mind that, eventually, all
the target space indices of the expression are going to be contracted
with the symmetric quantity
$\partial_{i_{1}}\ldots\partial_{i_{k}}\mathcal{W}(0)$. Therefore in
the calculation we only have to retain the symmetric part in the
indices and, at the same time, we can simplify expressions by
relabeling these indices. After a tedious calculation one arrives at
the result
\begin{eqnarray}
\sum_{i<j}\langle\Psi^{(i)}\Psi^{(j)}\rangle_{S} &=&
(-1)^{n}\sum_{i<j}^{n}\left[\hat{\psi}_{+}^{(i)}\hat{\psi}_{-}^{(j)}-
\hat{\psi}_{-}^{(i)}\hat{\psi}_{+}^{(j)}\right] \nonumber \\
&\times& {\partial^{2}\over \partial\hat{\varphi}^{(i)}
\partial\hat{\varphi}^{(j)}}
\int_{-{1\over 2}}^{1\over 2}d\xi\,\left(\hat{\varphi}^{(1)}+{\xi\over M}
\right)\ldots\left(\hat{\varphi}^{(n)}+{\xi\over M}\right),
\label{psipsi}
\end{eqnarray}
where the subscript $S$ indicates that we have retained only the symmetric
part. Plugging now (\ref{det}) and (\ref{psipsi}) back into (\ref{integrals})
and restoring the indices using (\ref{hats}) we arrive at Eq. 
(\ref{firstterm}).

In the case of the antiholomorphic superpotential the calculations
leading to (\ref{antihol}) are simpler than the ones presented above.
This is because $\bar{\sigma}^{\overline{\imath}}$ is independent of 
$\theta^{\pm}$ and then its Fourier transform is proportional to a 
fermionic delta function. This means that, for example, in a expression 
like
\begin{eqnarray}
\widetilde{
(\bar{K}{}^{\overline{\imath}_{1}}\star
\bar{\sigma}{}^{\overline{\imath}_{2}}\star
\ldots\star\bar{\sigma}
{}^{\overline{\imath}_{n}})}(\zeta)&=&
(-1)^{n-1}\int d^{2}\eta_{1}\ldots\int d^{2}\eta_{n-1}\,
\widetilde{\bar{K}}^{\,\overline{\imath}_{1}}(\eta_{1})
\widetilde{\bar{\sigma}}^{\,\overline{\imath}_{2}}(\eta_{2})\ldots
\widetilde{\bar{\sigma}}^{\,\overline{\imath}_{n}}(\zeta-\sum_{i=1}^{n-1}
\eta_{i})
\nonumber \\
&\times& \exp\left[{{1\over 2}\sum_{k<j}^{n-1}\langle\eta_{k},\eta_{j}\rangle}
\right]
\exp\left[{-{1\over 2}\sum_{i=1}^{n-1}\langle\zeta,\eta_{i}\rangle}\right].
\end{eqnarray}
Hence, $n-1$ integrations can be readily done and the whole integral is
reduced to a single integration. In the case of the term
containing
$\widetilde{\bar{\lambda}}_{+}{}^{\overline{i}_{1}}(\eta_{1})
\widetilde{\bar{\lambda}}_{-}{}^{\overline{i}_{2}}(\eta_{2})$,
$n-2$ integrations can be immediately done and one is left with
the calculation of two fermionic integrals. Keeping the term proportional
to $\zeta_{+}\zeta_{-}$ and restricting to the
symmetric part in the indices eliminates all dependence on the
deformation scale $M$ and the standard undeformed expression is obtained
for $\overline{\mathcal{W}} (\overline{\Phi})$.

Finally we outline the calculation of the deformed K\"ahler potential.
Again the terms $\bar{\sigma}^{\overline{\imath}}$ are mere spectators and the
relevant monomials to compute are\footnote{These monomials  
multiply $\overline{\varphi}{}^{\overline{\imath}_{2}}\ldots
\overline{\varphi}{}^{\overline{\imath}_{n}}$ and 
$\overline{\varphi}{}^{\overline{\imath}_{3}}\ldots
\overline{\varphi}{}^{\overline{\imath}_{n}}$ respectively.} 
\begin{eqnarray}
\widetilde{
({\Phi}^{i_{1}}\star\ldots\star{\Phi}^{i_{n}}\star
{\bar{K}}{}^{\,\overline{\imath}_{1}})}(\zeta)\Big|_{\zeta_{\pm}=0}, 
\hspace*{1cm}
\widetilde{({\Phi}^{i_{1}}\star\ldots\star\Phi^{i_{n}}\star
{\bar{\lambda}}_{+}{}^{\!\!\!\!\overline{\imath}_{1}} \star
{\bar{\lambda}}_{-}{}^{\!\!\!\!\overline{\imath}_{2}})}(\zeta)
\Big|_{\zeta_{\pm}=0}.
\label{monomials3}
\end{eqnarray}
Now we can apply the tricks used to calculate the holomorphic
superpotential. As a matter of example we comment on the first kind of
monomials where we can identify ${\bar{K}}{}^{\,\overline{\imath}}$
with an extra superfield. This reduces the problem to
Eq. (\ref{integrals}) with $n+1$ fields with the formal
identifications
\begin{eqnarray}
F^{(n+1)}\equiv
4\partial_{+}\partial_{-}\overline{\varphi}{}^{\,\overline{\imath}},
\hspace*{0.5cm}
{\varphi}^{(n+1)}\equiv {\overline{F}^{\,\overline{\imath}}
}, \hspace*{0.5cm}
{\psi}^{(n+1)}_{\pm}\equiv
\mp 2i{\partial_{\pm}\overline{\psi}_{+}{}^{\!\!\overline{\imath}}}.  
\end{eqnarray}
The calculation now follows the one outlined for the holomorphic
superpotential. For example, for the analog of the first term in
(\ref{integrals}) we find
\begin{eqnarray}
& &(-1)^{n}F^{(1)}\ldots F^{(n)}F^{(n+1)}\int
d^{2}\eta_{1}\ldots \int d^{2}\eta_{n}\,\, e^{\boldsymbol{\eta}_{+}^{T}
\mathcal{D}' \boldsymbol{\eta}_{-}} \nonumber \\
&=&\sum_{i=1}^{n}F^{(i)}{\partial\over \partial\varphi^{(i)}}
\int_{-{1\over 2}}^{1\over 2}d\xi\,\left(\varphi^{(1)}+{\xi\over M}
F^{(1)}\right)\ldots \left(\varphi^{(n)}+{\xi\over M}
F^{(n)}\right)\left(\varphi^{(n+1)}+{\xi\over M}
F^{(n+1)}\right) \nonumber \\
&+& F^{(n+1)}\int_{-{1\over 2}}^{1\over 2}d\xi\,
\left(\varphi^{(1)}+{\xi\over M}
F^{(1)}\right)\ldots \left(\varphi^{(n)}+{\xi\over M}
F^{(n)}\right),
\label{terms}
\end{eqnarray}
where by $\mathcal{D}'$ we have denoted the $n\times n$ analog of the 
matrix (\ref{matrix}).

We see already here the origin of the higher dimensional terms in 
Eq. (\ref{L2}). In the second line of (\ref{terms}) we find
two terms, one proportional to ${\varphi}^{(n+1)}=
\overline{F}^{\overline{\imath}}$ which will generate the term $F^{i}
\overline{F}^{\overline{\jmath}}$ in (\ref{L1}) and a second term
proportional to 
$F^{(n+1)}=4\partial_{+}\partial_{-}\overline{\varphi}^{\overline{\imath}}$
which is suppressed by $1/M$ and contains an extra power of $\xi$. This
gives rise to the first term on the right-hand side or (\ref{L2}).
On the other hand, the last line in Eq. (\ref{terms}) is at the origin of 
the first term in Eq. (\ref{L1}). Remember that Eq. (\ref{terms}) is
globally multiplied by $\overline{\varphi}{}^{\overline{\imath}_{2}}\ldots
\overline{\varphi}{}^{\overline{\imath}_{n}}$.  

The remaining terms of the K\"ahler potential come from the analog of
the second term inside the bracket in Eq. (\ref{integrals}) and the
ones coming from the second type of monomials in Eq. (\ref{monomials3}).
Again the calculations can be done mimicking the one for the superpotential.

\section*{Appendix C. The centrally extended superalgebra}
\renewcommand{\thesection}{C}
\setcounter{equation}{0}

In what follows we are going to discuss in more detail some aspects of
the centrally extended supersymmetry algebra
\begin{eqnarray}
Q_{\pm}^{2}=P_{\pm}, \hspace*{1cm} \{Q_{+},Q_{-}\}={1\over M}P_{+}P_{-}.
\label{Qalgebra}
\end{eqnarray}
In Section 3 we found that, in Euclidean space, this type of central
extension appears in nonanticommutative deformations (\ref{def}) of
(2,2) supersymmetic sigma models. Here, however, we are going to
forget about the restriction to Euclidean space and play the game of
studying the supersymmetry algebra (\ref{Qalgebra}) in Lorentzian
signature to discuss its consequence on the spectrum of a theory in 
Minkowski space in which this superalgebra would be realized.

Unitary representations of the two-dimensional Poincar\'e group are
labelled by the value of the Casimir operator $m\equiv
P_{+}P_{-}$. For $m=0$ the algebra reduces itself to the ordinary
(1,1) supersymmetry algebra. Therefore we focus our attention on the
representations with $m>0$. In the rest frame where $P_{\pm}=m$ and
after an obvious rescaling the algebra of supercharges can be written
as
\begin{eqnarray}
A^2=\mathbf{1}, \hspace*{1cm} B^2=\mathbf{1}, \hspace*{1cm}
\{A,B\}=\xi \mathbf{1},
\label{AB}
\end{eqnarray}
where $A$ and $B$ and hermitian operators and $\xi=m/M$.

The study of the unitary representations of the extended algebra
(\ref{AB}) proceeds by transforming it into an algebra of fermionic
creation-annihilation operators. These are defined by
\begin{eqnarray}
a=\alpha A+\beta B, \hspace*{1cm} a^{\dagger}=\alpha^{*} A+\beta^{*} B.
\end{eqnarray}
The complex constants $\alpha$, $\beta$ are fixed by demanding
$a^2=(a^{\dagger})^{2}=0$ and $\{a,a^{\dagger}\}=1$. Fixing the
overall phase ambiguity, one finds the solution
\begin{eqnarray}
\alpha={e^{i\gamma}\over 2\sin(2\gamma)}, \hspace*{1cm}
\beta={e^{-i\gamma}\over 2\sin(2\gamma)},
\end{eqnarray}
where $\gamma$ is defined by $\xi=-2\cos(2\gamma)$, which implies the BPS
condition
\begin{eqnarray}
|\xi|\leq 2 \hspace*{1cm} \Longrightarrow \hspace*{1cm} m\leq 2M.
\label{BPS}
\end{eqnarray}
Therefore unitary representations of the algebra exist provided the
theory does not contain states with masses above the cutoff
$\Lambda=2M$ given by Eq. (\ref{BPS}).  Below this bound unitary
irreducible representations contain two states, the fermionic vacuum
$|0\rangle$ and $|1\rangle\equiv a^{\dagger}|0\rangle$.

The previous analysis is valid whenever $|\xi|<2$. When the BPS bound
(\ref{BPS}) is saturated ($\xi=2$) the algebra reduces to $A^2=B^2=\mathbf{1}$,
$\{A,B\}=2$. Irreducible representations can then be constructed by
defining the operators $S_{\pm}=A\pm B$ which satisfy
\begin{eqnarray}
S_{+}^{2}=4,\hspace*{1cm} S_{-}^{2}=0.
\end{eqnarray}
Since $S_{\pm}^{\dagger}=S_{\pm}$, states which are $S_{-}$-exact have
zero norm and therefore should be removed from the spectrum to
preserve unitarity. As a result, irreducible representations are
one-dimensional and correspond to the short multiplets of
BPS-saturated states.

Here we have studied Hilbert space representations of the deformed
algebra (\ref{Qalgebra}) in two-dimensional Minkowski space-time.  A
very important question to be answered is, however, whether
two-dimensional quantum field theories exist in Lorentzian signature
in which this algebra of supercharges is realized. Our previous
analysis shows that finding such theories would be extremely
interesting since they would provide examples of quantum field
theories with a built-in cutoff.

\end{document}